\pdfoutput=1

\documentclass[11pt,twoside,a4paper,cmspaper,final,collab]{cms-tdr}
\def\svnVersion{d6c097c-D}\def\svnDate{2019/05/07}\def\cmsCernNoTag{CERN-EP-2018-278}\def\cmsCernDate{\today}\def\cmsMessage{"Published in Physics Letters B as \href{http://dx.doi.org/10.1016/j.physletb.2019.05.046}{\doi{10.1016/j.physletb.2019.05.046}.}"}

\begin{document}\cmsNoteHeader{EXO-17-015}

\hyphenation{had-ron-i-za-tion}
\hyphenation{cal-or-i-me-ter}
\hyphenation{de-vices}
\RCS$HeadURL$
\RCS$Id$

\newlength\cmsTabSkip\setlength{\cmsTabSkip}{1ex}
\newlength\cmsFigWidth
\ifthenelse{\boolean{cms@external}}{\setlength\cmsFigWidth{0.85\columnwidth}}{\setlength\cmsFigWidth{0.4\textwidth}}
\ifthenelse{\boolean{cms@external}}{\providecommand{\cmsLeft}{top\xspace}}{\providecommand{\cmsLeft}{left\xspace}}
\ifthenelse{\boolean{cms@external}}{\providecommand{\cmsRight}{bottom\xspace}}{\providecommand{\cmsRight}{right\xspace}}
\newcommand{\X}{\ensuremath{\mathrm{X}}\xspace}
\newcommand{\M}{\ensuremath{\mathrm{M}}\xspace}
\newcommand{\MLQ}{\ensuremath{{m_{\mathrm{LQ}}}}\xspace}
\newcommand{\Mmuj}{\ensuremath{{m_{\PGm\mathrm{j}}}}\xspace}
\newcommand{\dmx}{\ensuremath{\Delta_{\mathrm{\X,DM}}}\xspace}
\newcommand{\mdm}{\ensuremath{m_{\mathrm{DM}}}\xspace}
\newcommand{\mx}{\ensuremath{m_{\X}}\xspace}
\newcommand{\bzero}{\ensuremath{B}\xspace}
\newcommand{\wjets}{\ensuremath{\PW\text{+jets}}\xspace}
\newcommand{\zjets}{\ensuremath{\cPZ\text{+jets}}\xspace}
\newcommand{\yql}{\ensuremath{y_{\mathrm{Q}\ell}}\xspace}
\newcommand{\ydm}{\ensuremath{y_{\mathrm{D}}}\xspace}
\newcommand{\lqbf}{\ensuremath{{\mathcal{B}}(\mathrm{LQ}\to\cPqc\mu/\cPqs\mu)}\xspace}
\newcommand{\lqcbf}{\ensuremath{{\mathcal{B}}(\mathrm{LQ}\to\cPqc\mu)}\xspace}
\newcommand{\Msu}{\ensuremath{M_{\ensuremath{s}}^{\ensuremath{u}}}\xspace}
\newcommand{\Msd}{\ensuremath{M_{\ensuremath{s}}^{\ensuremath{d}}}\xspace}
\newcommand{\ylu}{\ensuremath{y_{\mathrm{Lu}}}\xspace}

\cmsNoteHeader{EXO-17-015}

\title{Search for dark matter in events with a leptoquark and missing transverse momentum in proton-proton collisions at 13\TeV}

\date{\today}

\abstract{
A search is presented for dark matter in proton-proton collisions
at a center-of-mass energy of $\sqrt{s}=13\TeV$ using events
with at least one high transverse momentum (\pt) muon,
at least one high-\pt jet, and large missing transverse momentum.
The data were collected with the CMS detector at the CERN LHC
in 2016 and 2017,
and correspond to an integrated luminosity of 77.4\fbinv.
In the examined scenario, a pair of scalar leptoquarks
is assumed to be produced.
One leptoquark decays to a muon and a jet
while the other decays to dark matter
and low-\pt standard model particles.
The signature for signal events would be significant missing
transverse momentum from the dark matter
in conjunction with a peak at the leptoquark mass in the
invariant mass distribution of the highest \pt muon and jet.
The data are observed to be consistent with the background
predicted by the standard model.
For the first benchmark scenario considered, dark matter masses up to 500\GeV
are excluded for leptoquark masses $\MLQ\approx1400\GeV$,
and up to 300\GeV for $\MLQ\approx1500\GeV$.	
For the second benchmark scenario, dark matter masses up to 600\GeV are excluded for $\MLQ\approx 1400\GeV$.
}

\hypersetup{%
pdfauthor={CMS Collaboration},%
pdftitle={Search for dark matter in final states with a muon, a jet, and missing transverse momentum in proton-proton collisions at 13 TeV},%
pdfsubject={CMS},%
pdfkeywords={CMS, physics, dark matter, leptoquarks}}

\maketitle

\section{Introduction}
\label{sec:introduction}

Dark matter (DM) has been a subject of intense interest for decades.
Extensive astrophysical evidence for DM
exists~\cite{Bertone:2004pz,Feng:2010gw,Porter:2011nv},
such as from observations of the dynamics of galaxy clusters
and measurements of anisotropies in the cosmic microwave background.
Nonetheless,
the nature of DM remains unknown and it has not been
observed outside the astrophysical context.
Its relic density is determined to be
$\Omega_{\text{DM}}= (0.1186\pm 0.0020)/h^2$~\cite{PDG2018,Ade:2015xua},
where $h$ is the Hubble constant.

Dark matter could potentially be created
in high-energy proton-proton ($\Pp\Pp$) collisions,
such as at the CERN LHC.
Because of its presumed weakly interacting nature,
a DM particle produced
at the LHC would escape unobserved,
manifesting itself as missing transverse momentum \ptvecmiss
in the reconstructed events.
The most generic signal for DM at the LHC thus consists of an excess,
relative to the standard model (SM) expectation,
of events with sizable \ptvecmiss recoiling against a visible SM object
such as a jet, a photon, or an electroweak boson.
Such searches have been conducted at the LHC by the ATLAS and CMS
Collaborations~\cite{Sirunyan:2017jix,Sirunyan:2017qfc,Sirunyan:2017ewk,Aaboud:2017dor,Aaboud:2016tnv,Aaboud:2016qgg,Aaboud:2017phn,Aaboud:2018xdl,Sirunyan:2018fpy}
but with no evidence, to date, for DM~\cite{Penning:2017tmb}.
The absence of a signal in these generic searches
suggests that alternative strategies should be pursued.

In this Letter,
we present a search for DM at the LHC
using a new approach,
based on the coannihilation
paradigm
introduced in Ref.~\cite{Baker:2015qna}.
The data,
corresponding to an integrated luminosity of 77.4\fbinv
of $\Pp\Pp$ collisions,
were collected by the CMS Collaboration in 2016 (35.9\fbinv)
and 2017 (41.5\fbinv)
at a center-of-mass energy of $\sqrt{s}=13\TeV$.
The coannihilation process considered arises within
a general class of simplified models in which a DM particle is either
annihilated or produced in conjunction
with a so-called coannihilation partner, denoted ``\X''.
At the LHC,
DM could thus be produced
through a reaction like $\M\to\text{DM} + \X$,
where \M is a mediator representing either
an SM or a beyond-the-SM particle.
To be consistent with the observed value of $\Omega_{\text{DM}}$,
the fractional mass difference $\dmx\equiv(\mx - \mdm)/\mdm$
between the \X and DM particles should be
less than $\approx$0.2~\cite{Baker:2015qna}.

The considered coannihilation paradigm
introduces many DM signatures
that are not covered by current searches.
Here,
we consider the principal case-study scenario of Ref.~\cite{Baker:2015qna},
in which the mediator \M is a scalar leptoquark (LQ) doublet
and the particle \X is a new Dirac fermion.
An LQ is a hypothetical color-triplet, fractionally charged
boson that carries both lepton and baryon quantum numbers.
Leptoquarks appear in many extensions of the SM,
such as grand unification
theories~\cite{Pati:1973uk,Pati:1974yy,GUT,Fritzsch:1974nn}
and models with composite quarks and leptons~\cite{COMP1}.
To be consistent with experimental constraints
on flavor changing neutral currents,
we assume the LQ to couple to a single SM flavor generation
only~\cite{Dorsner:2016wpm},
taken to be the second generation for this study.
We choose the second generation because muons provide
a clear experimental signature.
We further assume pair production of LQs,
as is predominantly expected in $\Pp\Pp$ collisions.
Following Ref.~\cite{Baker:2015qna},
the DM particle is assumed to be a Majorana fermion
with the gauge group structure (1, 1, 0),
where the numbers in parentheses indicate the color
SU(3)$_{\mathrm{C}}$,
weak isospin SU(2)$_{\mathrm{L}}$,
and weak hypercharge U(1)$_{\mathrm{Y}}$
multiplet dimensions,
respectively.
We use the convention
$Q=T_3+Y/2$ for the electric charge $Q$ of the particle,
with $T_3$ the third component of weak isospin and Y the weak hypercharge.
The corresponding assignments for both the \X and LQ particles
are (3, 2, 7/3).
For the LQ,
$T_3=\pm1/2$ and $Q=5/3,2/3$ for the two members of the doublet, respectively.
The interaction term of the Lagrangian for this model is:
 \begin{linenomath}
\begin{equation}
\mathcal{L}=-(\ydm\overline{\mathrm{X}}M_{s}\:\mathrm{DM}+\yql\overline{Q_L}M_{s}\ell_R+\ylu\overline{L_L}M_{s}^cu_R+\text{h.c.})
\end{equation}
\end{linenomath}
where the superscript c refers to charge conjugation. After the breaking of electroweak symmetry, the different doublet components of X, $M_s$, $Q_L$, and $L_L$ take the form: 
 \begin{linenomath}
\begin{equation}
\mathcal{L}=-\ydm\overline{\mathrm{X}}^{u}M_{s}^{u}\:\mathrm{DM}-\yql\overline{u}_L M_{s}^{u}\ell_R-\ylu\overline{e}_L(M_{s}^{u})^c u_R 
\end{equation}
\end{linenomath} 
 \begin{linenomath}
\begin{equation}
-\ydm\overline{\mathrm{X}}^{d}M_{s}^{d}\:\mathrm{DM}-\yql\overline{d}_L M_{s}^{d}\ell_R-\ylu\overline{\nu}_L(M_{s}^{d})^c u_R 
\end{equation}
\end{linenomath} 

\begin{figure}[tbp]
\centering
\includegraphics[width=0.49\textwidth]{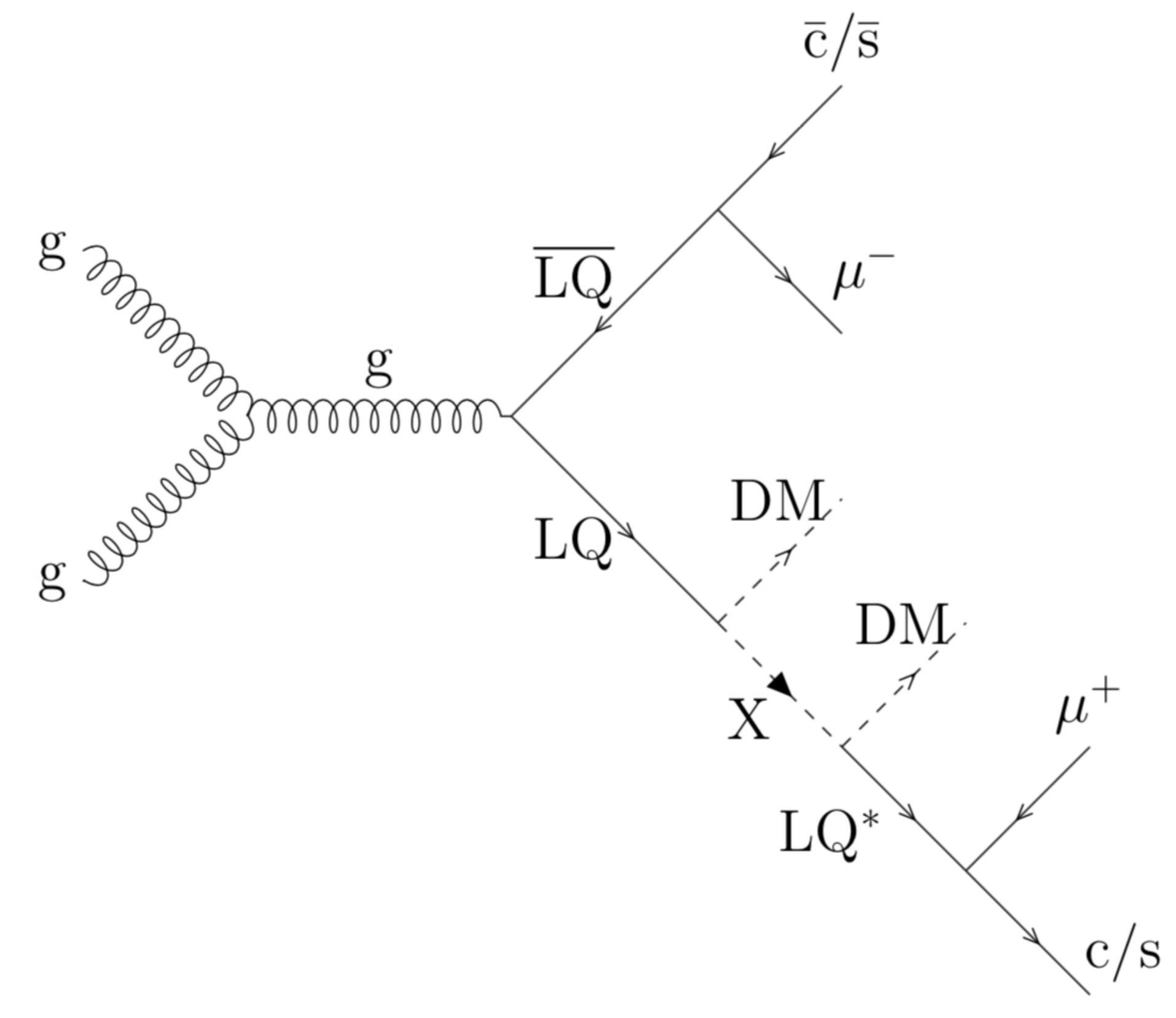}
\caption{
An example Feynman diagram for the signal process considered in this study,
where \Pg is a gluon, LQ a leptoquark, DM a dark matter particle,
and \X a new Dirac fermion.
The superscript ``*'' indicates an off-shell particle.
}
\label{fig:feynman}
\end{figure}

The first line represents the interactions of the upper component of the LQ doublet \Msu, which has electric charge -5/3. The first term in the first line describes the decay of the LQ to DM and X, with \ydm the coupling strength. The second and third terms of the first line represent the interactions, with coupling strengths \yql and \ylu, for the different helicity couplings to the up-type quark and the lepton. The second line describes the interactions of the lower component of the leptoquark doublet \Msd, which has electric charge -2/3.  In the limit that $\yql \neq 0$ and $\ylu = 0$, both the upper and lower components of the doublet have the same collider phenomenology. The SM decays of the upper component of the \Msu doublet are to up-type quarks and leptons, while those of the lower component of the \Msd doublet are to down-type quarks and leptons. In the limit that $\yql = 0$ and $\ylu \neq 0$, the upper and lower components of the doublet have different collider phenomenology. The SM decays of the upper component of \Msu are again to up-type quarks and leptons, but the only SM decays of the lower component are to neutrinos and up-type quarks, and not to the full leptoquark signature. Both the upper and lower components of the LQ doublet have been considered in this analysis. Following the example of Ref.~\cite{Baker:2015qna}, we assume $\ylu = 0$. Thus it should be born in mind that the limits obtained in this analysis are valid under this explicit assumption.
In the pair production of the LQ, one LQ decays to a muon and a \cPqc\ or an \cPqs\ quark,
while the other decays
through the coannihilation paradigm,
to DM and~\X.
The \X particle subsequently decays through a crossed coannihilation
process to a DM particle and an off-shell LQ,
where the decay products of this latter particle,
also a muon and a \cPqc\ or an \cPqs\ quark,
have low transverse momentum (\pt)
because of the smallness of \dmx
and are potentially undetected.
An example Feynman diagram is shown in Fig.~\ref{fig:feynman}.

The most restrictive lower limit on the mass of a pair-produced second-generation LQ, assuming an LQ branching fraction $\lqcbf=100\%$, is currently 1530\GeV~\cite{Sirunyan:2018ryt}.
However, when the decay to DM and \X is allowed,
the $\mathrm{LQ}\to\cPqc\mu/\cPqs\mu$ branching fraction is reduced
and the limit on \MLQ becomes weaker.
The branching fraction then also depends on \mdm, $\dmx$, and \bzero, where \bzero is defined as \lqbf in the limit of massless \X and DM particles and is related to
\yql and \ydm  by the following formula:
\begin{linenomath}
\begin{equation}
  \bzero = \left.\lqbf\right|_{\mdm=\mx=0}
     = \frac{\yql^2}{\yql^2 + 2\ydm^2}.
\end{equation}
\end{linenomath}
Following Ref.~\cite{Baker:2015qna},
we set $\ydm=0.1$ and $\dmx=0.1$.
We consider two values for \bzero: 0.5 and~0.1.
Recasting the results of Ref.~\cite{Sirunyan:2018ryt}
for the upper component of the LQ doublet,
and taking $\mdm=300\GeV$ as a representative value,
the lower limit on a second-generation LQ
is reduced to 1340\GeV for $\bzero=0.5$
and to 960\GeV for $\bzero=0.1$.

In this analysis, the final state consists of
a high-\pt muon and a high-\pt jet
from the decay of the on-shell LQ,
\ptvecmiss from the DM particles,
and low-\pt SM objects from the decay of the off-shell LQ.
Note that,
in the analysis,
we do not employ \cPqc\ quark tagging criteria
but rather---to improve the signal event selection efficiency---utilize
generic untagged jets as the \cPqc\ quark jet candidates.
The existence of the signal process is inferred by
a peak at the LQ mass \MLQ,
in the invariant mass \Mmuj distribution
of the high-\pt muon and jet,
in conjunction with significant \ptvecmiss from the DM.
This peak at the LQ mass provides a striking experimental signature in the search
for signal processes containing DM.
In contrast,
generic searches for DM,
in which there are no new particles other than DM
and intermediate mediator states,
mostly rely on
a mere enhancement in the tail of the \ptvecmiss distribution.

The principal SM backgrounds in this search arise
from events with a {\PW} boson and jets
(\wjets) or with a top quark-antiquark (\ttbar) pair:
in both cases,
the leptonic decay of a {\PW} boson can yield
a high-\pt muon and neutrino,
where the neutrino can lead to significant \ptvecmiss.
Events with single top quark or diboson ($\PW\PW$, $\PW\PZ$, and $\cPZ\cPZ$)
production similarly can enter the background,
although at a lower level.
Other smaller sources of SM background arise from
quantum chromodynamics (QCD) events,
namely events with a multijet final state produced exclusively
through the strong interaction,
and from events with a {\cPZ} boson and jets (\zjets).
A QCD event can enter the background if a muon and a neutrino are
produced through the semileptonic decay of a quark,
or if a jet is erroneously identified as a muon
in conjunction with spurious \ptvecmiss arising
from the mismeasurement of jet~\pt.
Events with \zjets production can enter the background if one
of the leptons in $\cPZ\to\MM$ decays is not reconstructed
or lies outside the acceptance of the detector,
leading to \ptvecmiss,
or if \ptvecmiss arises because of misreconstructed jet~\pt.

\section{The CMS detector and trigger}
\label{sec:detector}

The central feature of the CMS detector is a superconducting
solenoid of 6\unit{m} internal diameter, providing a
magnetic field of 3.8\unit{T}. Within the solenoid volume
are a silicon pixel and strip inner tracker,
a lead tungstate crystal electromagnetic calorimeter,
and a brass and scintillator hadron calorimeter,
each composed of a barrel and two endcap sections. Extensive forward
calorimetry complements the coverage provided by the barrel
and endcap detectors. Muons are
detected in gas-ionization chambers embedded in the steel
flux-return yoke outside the solenoid.
A detailed description of the CMS detector, together with a
definition of the coordinate system and the relevant
kinematic variables, is given in Ref.~\cite{Chatrchyan:2008zzk}.

Events of interest are selected using a two-tiered
trigger system~\cite{Khachatryan:2016bia}.
The first level, composed of custom hardware processors,
uses information from the calorimeters and muon detectors to
select events at a rate of around 100\unit{kHz}.
The second level,
based on an array of microprocessors
running a version of the full event reconstruction software
optimized for fast processing,
reduces the event rate to around 1\unit{kHz} before data storage.
The set of triggers used for this analysis requires events
to contain a muon with $\pt>50$\GeV.

\section{Event reconstruction and selection}
\label{sec:reconstruction}

Individual particles are reconstructed with the CMS
particle-flow (PF) algorithm~\cite{Sirunyan:2017ulk},
which identifies them as charged hadrons,
neutral hadrons, muons, electrons, or photons.
Muon reconstruction is performed by
matching a track segment reconstructed in the inner tracker
with a track segment reconstructed in the muon detector and
performing a global fit of the hits
from the two track segments.
The candidate muons are required to satisfy
the tight selection criteria of Ref.~\cite{Sirunyan:2018fpa},
to pass within 2\unit{mm} of the primary event vertex
in the direction along the beam axis and within 0.45\unit{mm}
in the plane perpendicular to that axis,
and to have $\pt>60\GeV$ and $\abs{\eta}<2.4$,
where $\eta$ is the pseudorapidity.
Electrons are reconstructed by matching energy deposits in the
electromagnetic calorimeter with
track segments in the inner tracker~\cite{Khachatryan:2015hwa}
and are required to have $\pt>15\GeV$ and $\abs{\eta}< 2.5$.
Hadronically decaying $\tau$ leptons (\tauh)
are reconstructed using the
hadrons-plus-strips algorithm described in Ref.~\cite{Sirunyan:2018pgf}
and are required to have $\pt>20\GeV$ and $\abs{\eta}< 2.3$.
The electron and \tauh candidates are mainly used to veto \zjets and diboson events, as described below, and are selected using loose~\cite{Khachatryan:2015hwa,Sirunyan:2018pgf} identification criteria.

The primary event vertex is defined to be the reconstructed interaction vertex
with the largest value of summed physics-object $\pt^2$.
The physics objects considered for this purpose are the jets found by
clustering the charged-particle tracks assigned to the vertex,
using the anti-\kt jet finding algorithm~\cite{Cacciari:2008gp,Cacciari:2011ma}
with a distance parameter~0.4,
and the associated missing transverse momentum,
taken as the negative of the vector \pt sum of those jets.
The transverse momentum imbalance \ptvecmiss in an event is
calculated as the negative of the vector \pt sum of all PF candidates.
Its magnitude is denoted~\ptmiss.
Events are required to have $\ptmiss>100\GeV$.

To suppress the contributions of muons that arise from hadron decays,
muon candidates are subjected to an isolation requirement.
The scalar \pt sum of charged hadron, neutral hadron, and
photon PF candidates within a cone of radius
$\Delta R \equiv$ $\sqrt{\smash[b]{(\Delta\eta)^2 + (\Delta\phi)^2}} = 0.4$
around the muon direction,
where $\phi$ is the azimuthal angle,
is calculated.
The expected contributions of neutral particles
from additional \Pp\Pp\ interactions in the same or neighboring bunch crossings
(pileup) are subtracted~\cite{Cacciari:2007fd}.
An isolation variable $I$ is defined by dividing this sum
by the muon \pt.
The isolation requirement is $I<0.15$.
At least one isolated muon candidate is required to be present in the event.

The reconstruction of jets is performed by clustering PF candidates
using the anti-\kt algorithm with a distance parameter of~0.4,
excluding charged-particle tracks not associated with the primary vertex.
The jet energies are corrected for the combined
response function of the calorimeters~\cite{Khachatryan:2016kdb}
and to account for the expected contributions of
neutral particles from pileup~\cite{CMS-PAS-JME-14-001,Cacciari:2007fd}.
Jets are required to appear within $\abs{\eta}<2.4$.
Bottom (\cPqb) quark jets are identified (\cPqb\ tagged) from this sample
using the combined secondary vertex (CSVv2)
algorithm at the tight working point~\cite{Sirunyan:2017ezt},
which yields a \cPqb\ quark jet identification efficiency of approximately 40\%, and a misidentification probability of about 0.1\% for gluon and light-flavored quark jets and of about 2\% for charm quark jets.
Jets tagged as {\cPqb} jets are required to have $\pt>30\GeV$.

The leading (highest \pt) jet in an event is required to
have $\pt>100\GeV$ and to be separated by $\Delta R>0.5$ from the leading isolated muon candidate. The leading isolated muon and leading jet are then
combined to form the LQ candidate.
Studies with simulated signal events establish that,
for the values of model parameters used in the present study,
this matching identifies the correct combination
over 98\% of the time.

To suppress background from \ttbar production,
events are rejected if they contain a {\cPqb}-tagged jet,
an electron candidate, or a \tauh candidate.
The veto on events with a {\cPqb}-tagged jet reduces the
background from events with a top quark by more than a factor of 2
while reducing the signal efficiency by only around 10\%.
The vetoes on electron and \tauh candidates also suppress
the \wjets and \zjets background.
The \wjets background is further suppressed by requiring
the transverse mass \mT~\cite{Arnison:1983rp} formed from
the \pt vector of the leading muon and \ptvecmiss to exceed 100\GeV.
The \zjets background is further reduced
by eliminating events
with a loosely identified and isolated ($I<0.25$) muon candidate
if that muon candidate has $\pt>10\GeV$,
an opposite charge to the leading muon,
and forms a dimuon mass with the leading muon
within 10\GeV of the {\cPZ} boson mass.

Background from QCD events mostly arises when the \pt of one of
the highest \pt jets is underestimated or when a hadron in a jet
undergoes a semileptonic decay,
introducing \ptvecmiss that is aligned with that jet.
To suppress this background,
the angular difference $\Delta\phi$ between the leading jet
and \ptvecmiss,
and between the leading muon and \ptvecmiss,
is required to exceed~0.5.

The above requirements are referred to as the ``preselection" criteria,
and form the basis for the definition
of several control regions used to evaluate background,
as described in Section~\ref{sec:background}.
The final selection criteria,
corresponding to the signal region,
are the same as the preselection criteria
except for a more stringent requirement on \mT, $\mT>500\GeV$.
This condition is determined from optimization
studies utilizing simulated signal and SM event samples.
For $\MLQ>800\GeV$,
the signal efficiency for events satisfying the preselection
criteria is around 73\%,
essentially independent of~\MLQ.
For the signal region criteria,
the signal efficiency varies from 47 to 63\%
as \MLQ increases from 800 to 1500\GeV.

\section{Event simulation}
\label{sec:simulation}

Monte Carlo (MC) simulation of signal and background processes
is used to validate the analysis procedures,
evaluate background,
and determine the signal efficiency.
Simulation of \ttbar and single top quark events
is performed at next-to-leading-order (NLO) accuracy with the
\POWHEG 2.0~\cite{Frixione:2007vw,Nason:2004rx,Alioli:2010xd,Alioli:2009je,Re:2010bp,Frixione:2007nw}
event generator.
To describe \wjets and \zjets production,
the \MGvATNLO 2.2.2~\cite{Alwall:2014hca,Alwall:2007fs} program
at leading order (LO) is used.
A $K$ factor,
calculated as described in Ref.~\cite{Sirunyan:2017jix},
is applied as a function of boson \pt to account for next-to-NLO (NNLO) corrections.
The statistical precision of our available \MGvATNLO \wjets sample
is low for off-shell \PW\ boson masses above around 100\GeV.
Therefore, to describe \wjets production for \PW\ boson masses above 100\GeV,
we use the LO \PYTHIA 8.212~\cite{Sjostrand:8.2} program with the
CUETP8M1 tune~\cite{Khachatryan:2015pea},
rather than \MGvATNLO,
with a $K$ factor to account for NNLO corrections
applied as a function of the \PW\ boson mass.
This $K$ factor is determined using the
\FEWZ 3.1~\cite{Gavin:2010az, Li:2012wna} program.
Diboson production is simulated at NLO using either the
\POWHEG~\cite{Melia:2011tj}
or \MGvATNLO 2.2.2 generators.

Signal events are simulated at LO using \MGvATNLO 2.2.2.
The signal samples are generated for LQ and DM masses in the
ranges $800\leqslant\MLQ\leqslant1500\GeV$ in 100\GeV steps and $300\leqslant\mdm\leqslant700\GeV$ in 50\GeV steps, respectively,
with cross sections normalized to
NLO~\cite{Kramer:2004df,Kramer:1997hh}
accuracy.

For simulated samples at LO (NLO),
the NNPDF3.0LO (NNPDF3.0NLO)~\cite{Ball:2014uwa}
parton distribution functions (PDFs) are used.
All samples are interfaced to \PYTHIA 8.212
to describe parton showering and hadronization.
Pileup interactions are modeled using simulated
minimum bias event samples generated with \PYTHIA,
with the distribution of $\Pp\Pp$ interactions
per bunch crossing adjusted to reproduce the observed spectrum.

The response of the CMS detector is modeled,
for both the signal and background samples,
using the \GEANTfour\cite{Agostinelli:2002hh}
suite of programs.
Small differences between the data and simulation in the trigger,
particle identification, and muon isolation efficiencies,
and in the jet \pt resolution and \ptmiss,
are accounted for through the application of scale factor corrections.

\section{Background estimation}
\label{sec:background}

\begin{figure}[htb!]
\centering
\includegraphics[width=0.49\textwidth]{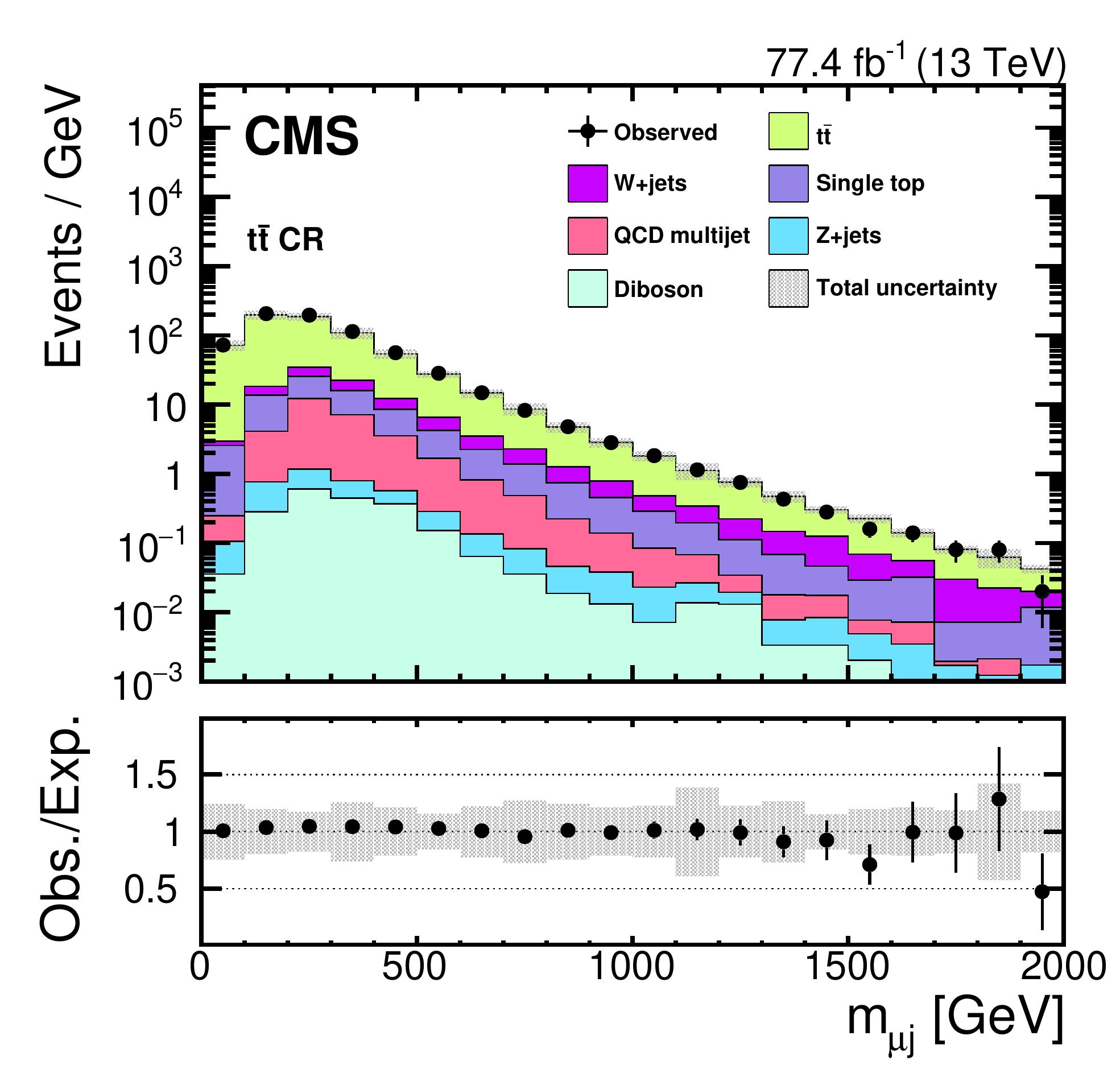}
\includegraphics[width=0.49\textwidth]{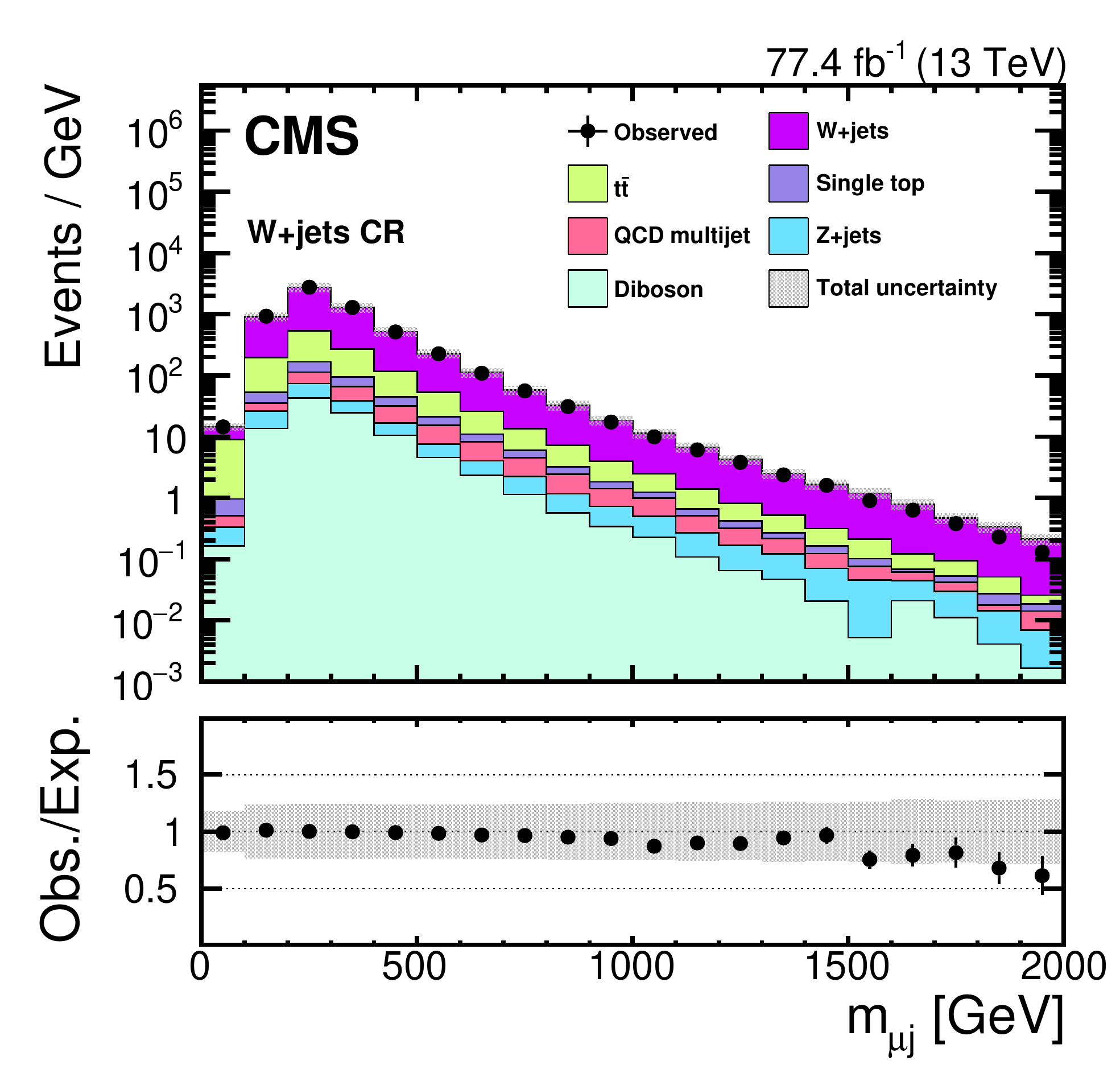}
\caption{
The \Mmuj distributions in data and simulation for the
(\cmsLeft) \ttbar-
and (\cmsRight) \wjets-enriched
control samples for the combined 2016 and 2017 data sets.
The respective data-to-simulation normalization scale factors
have been applied to the simulated distributions.
The lower panels show the ratio of the observed to the simulated results.
The vertical error bars on the data points are statistical.
The gray band shows the total uncertainty in the background prediction,
including both statistical and systematic terms.
}
\label{fig:TTbarCR}
\end{figure}

Background from \ttbar production is evaluated from simulation,
with the events reweighted to reproduce the observed
distribution of the top quark \pt~\cite{Khachatryan:2015oqa,Chatrchyan:2012saa}.
The normalization is determined using a data-to-simulation scale factor
derived from a \ttbar-enhanced control sample.
The control sample is defined using the preselection criteria
of Section~\ref{sec:reconstruction}
except that events are required to contain at least one {\cPqb}-tagged jet.
The purity of \ttbar events in the control sample is estimated
to be 85\%.
The remaining 15\% of the sample is composed primarily
of events with \wjets and single top quark production.
The scale factor is defined as the ratio of the
number of events in the control sample from data,
after subtraction of the non-\ttbar components, estimated from simulation,
to the number in the corresponding simulated \ttbar control sample
scaled to the same integrated luminosity.
The scale factor is found to be $0.95\pm0.01\stat$ for
the 2016 data and $1.16\pm0.01\stat$ for the 2017 data
and is essentially independent of the \mT selection requirement.
The difference in the scale factors
between 2016 and 2017 arises from changes in the running conditions,
the reconstruction procedures,
and the tuning of the simulation programs.
Figure~\ref{fig:TTbarCR} (\cmsLeft) shows the distribution of \Mmuj
in the combined 2016+2017 \ttbar control sample for data and simulation.

The systematic uncertainty in the scale factor is derived using
an orthogonal {\ttbar}-enhanced control sample,
selected with the preselection criteria
except that events must contain at least one electron candidate,
in addition to the existing muon candidate,
and the veto on the presence of a {\cPqb}-tagged jet is removed.
The purity of \ttbar events in this control sample is around 90\%,
with the remainder of the events arising primarily
from single top quark production.
The scale factor obtained is $0.85\pm0.01\stat$ for
the 2016 data and $1.00\pm0.01\stat$ for the 2017 data.
On the basis of these results,
a systematic uncertainty of 10\% is assigned to
both the 2016 and 2017 scale factors
discussed in the previous paragraph.

The background from \wjets production
is similarly estimated from simulation,
with a correction to the normalization obtained from
a \wjets-dominated control sample.
The control sample is defined using the preselection criteria
except with the additional requirement $50<\mT<150\GeV$.
The purity of \wjets events in this sample is about 80\%.
The main remaining contribution is from \ttbar production.
A data-to-simulation normalization scale factor is
obtained by subtracting the non-\wjets contribution,
estimated from simulation,
from the control sample in data,
and dividing the resulting number of events by
the number of events in the simulated control region
normalized to the same integrated luminosity.
The \ttbar scale factor has been applied to the \ttbar background before subtracting it from the control sample in data.
The scale factor is found to be
$1.02\pm0.01\stat$ and
$1.11\pm0.01\stat$ for the 2016 and 2017 data, respectively.
The distribution of \Mmuj in the combined 2016+2017 \wjets control sample,
for data and simulation,
is shown in Fig.~\ref{fig:TTbarCR} (\cmsRight).

To verify the stability of the scale factor in regions with larger \mT, corresponding to off-shell {\PW} boson production, the data-to-simulation scale factor has been remeasured in control regions with \mT in the ranges 150 to 200, 200 to 300 and 300 to 400\GeV. The measured scale factor is in agreement within 20\% with the original scale factor. In a second test, we examine the level of agreement between the data and simulation for the normalization of the \Mmuj distribution in a {\cPZ}($\to\MM$)+jets control sample with one of the two muons in each event removed to emulate a sample of \wjets events. This control sample is selected by requiring two oppositely charged isolated muons with $\pt>30\GeV$, with a dimuon invariant mass between 80 and 100\GeV, but with otherwise similar selection criteria to those of the preselection. The {\cPZ} boson candidate is boosted to its rest frame, the dimuon mass is scaled to the mass of the {\PW} boson, and the system is then boosted back to the original laboratory frame. One of the two muons is randomly removed to simulate a neutrino from {\PW} boson decay. The resulting missing momentum is added to the \ptvecmiss of the event and the value of \mT recalculated before applying the signal region selection criteria. The level of agreement in the resulting distribution of \Mmuj between data and simulation is also around 20\%. On the basis of these two studies, a 20\% uncertainty is assigned to the \wjets background prediction.
In addition, relevant theoretical uncertainties are also taken into consideration in the shape and normalization of the \wjets background in the signal regions. This is discussed in Section~\ref{sec:systematics}.

The background from QCD processes is expected to be small.
However,
since the QCD background primarily arises as
a consequence of jet mismeasurement,
it is not well modeled by simulation.
Thus,
we evaluate the QCD background using a
method based primarily on data.
A control sample is defined using the signal region
criteria of Section~\ref{sec:reconstruction}
except that muon candidates are required to
fail the isolation condition.
To estimate the QCD background in the signal region,
the events in the control sample are weighted as a function of muon \pt by
a muon misidentification probability,
called the jet-to-muon misidentification rate,
determined in a QCD-enriched event sample
denoted the ``low-$\Delta\phi$'' sample.
The low-$\Delta\phi$ sample is defined in
the same manner as the signal region
except the angle $\Delta\phi$ between \ptvecmiss and the leading jet
is required to be $\Delta\phi<0.5$ rather than $\Delta\phi>0.5$
and we require $\mT>100\GeV$ rather than $\mT>500\GeV$.
The jet-to-muon misidentification rate is defined,
from this sample,
as the ratio of the number of events
that satisfy the muon relative isolation criterion $I<0.15$
to the number of events with no requirement on the isolation,
after subtracting the non-QCD components,
evaluated with simulation,
from both the numerator and denominator.
The purity of QCD events in the numerator is about 25\%,
while that in the denominator is approximately 70\%.
The jet-to-muon misidentification rate is parameterized in terms of the muon \pt
using an analytical function
and varies from 5\% for muon $\pt=60\GeV$ to 50\% for $\pt>300\GeV$.

A 50\% uncertainty is assigned to the QCD background prediction
to account for the uncertainty in the jet-to-muon misidentification rate.
This uncertainty primarily arises from the uncertainties in the
normalization of the non-QCD processes subtracted from the numerator
and denominator in calculating the misidentification rate.
Other sources of uncertainty,
such as the choice of the analytic function used
to parameterize the jet-to-muon misidentification rate,
or the uncertainties in the values of the fit parameters,
are negligible in comparison.

The backgrounds from events with diboson, single top quark, and \zjets
production are estimated from simulation.
An uncertainty of 15\% is assigned to the diboson background prediction
based on the level of agreement between data and simulation
in a diboson-enhanced control sample
selected by requiring events to contain three leptons
(\Pe\ or {\PGm}),
two of which are consistent with arising from {\cPZ} boson decay.
Uncertainties of 15 and 10\% are assigned
to the single top quark and \zjets backgrounds,
respectively,
based on the results of Refs.~\cite{Sirunyan:2016cdg}
and~\cite{Sirunyan:2018cpw}.

\section{Systematic uncertainties}
\label{sec:systematics}

We evaluate systematic uncertainties that affect
the normalization or shape of the \Mmuj spectrum,
either in the signal or background predictions.
Uncertainties specific to individual background components
were presented in Section~\ref{sec:background}.

The uncertainty in the trigger efficiency is 5\%.
Those in the muon reconstruction and isolation
efficiency~\cite{tag-and-probe},
the electron reconstruction efficiency~\cite{tag-and-probe},
and the \tauh reconstruction efficiency~\cite{TAU-11-001,Khachatryan:2015dfa}
are 5, 2, and 2\%, respectively.
The uncertainty related to the \cPqb-tagging misidentification
is~1\%~\cite{Sirunyan:2017ezt}.
The uncertainty in the pileup description in simulation is assessed by varying
the total inelastic cross section by 4.6\%~\cite{Sirunyan:2018nqx},
and is found to be 1\%.
Statistical uncertainties related to
the limited number of events in the data control samples
are accounted for as described in Ref.~\cite{BarlowBeeston},
while those related to the limited number of events in simulation
are accounted for by allowing the content in each bin of the simulated
distributions to vary within its statistical uncertainty.
The uncertainty in the integrated luminosity
is 2.5\%~\cite{CMS-PAS-LUM-17-001} for the 2016 data
and 2.3\%~\cite{CMS-PAS-LUM-17-004} for the 2017 data.
Uncertainties associated with the jet energy scale,
the jet energy resolution,
and \ptmiss are also evaluated~\cite{CMS-PAS-JME-16-003,CMS-PAS-JME-16-004}.
These uncertainties affect both the shape and normalization
of the simulated signal and background distributions.
Uncertainties related to the top quark \pt reweighting in simulated \ttbar events
are evaluated by varying the reweighting parameters between zero
and twice their nominal values~\cite{Khachatryan:2015oqa,Chatrchyan:2012saa}.
Uncertainties related to the PDFs, evaluated for the signal acceptance, are determined following the PDF4LHC prescription~\cite{Butterworth:2015oua} and are found to be 3\%. Those related to the renormalization and factorization scales, evaluated for the signal yields and for the \ttbar and \wjets backgrounds, are estimated by varying each scale independently, and also coherently, by a factor of 2.0 and~0.5. The largest variations upward and downward in the results are used to define an uncertainty envelope. This uncertainty amounts to 1\% for signal events, independent of the LQ mass. For the \ttbar and \wjets backgrounds, this uncertainty varies between 5 and 15\% depending on \MLQ and thus accounts for a systematic uncertainty in the shape of the distribution.
The uncertainty associated with the
method chosen to determine the $K$ factor for the \wjets
background evaluation is estimated to be~5\%.

\begin{table*}[tbh]
\topcaption{
Systematic uncertainties affecting the normalization
of signal and background distributions. The PDF uncertainty affects the signal distribution only, while the other uncertainties affect both the signal and background distributions.
}
\centering
\begin{tabular}{lc}
\hline
Item & Relative uncertainty (\%) \\
\hline
Trigger efficiency                              & 5.0 \\
Muon identification efficiency                  & 5.0 \\
Electron identification (veto) efficiency       & 2.0 \\
\tauh lepton identification (veto) efficiency   & 2.0 \\
{\cPqb} jet identification (veto) efficiency    & 1.0 \\
Pileup modeling                                 & 1.0 \\
Integrated luminosity                           & 2.5 (2016), 2.3 (2017)\\
PDF                                             & 3.0 \\
Renormalization and factorization scales        & 1.0 \\
\ttbar normalization    & 10 \\
W+jets normalization    & 20 \\
\hline
\end{tabular}
\label{tab:systematics}
\end{table*}

Table~\ref{tab:systematics} summarizes the
systematic uncertainties affecting the normalization of distributions.

\section{Results}
\label{sec:results}

\begin{figure}[tbh]
\centering
\includegraphics[width=0.49\textwidth]{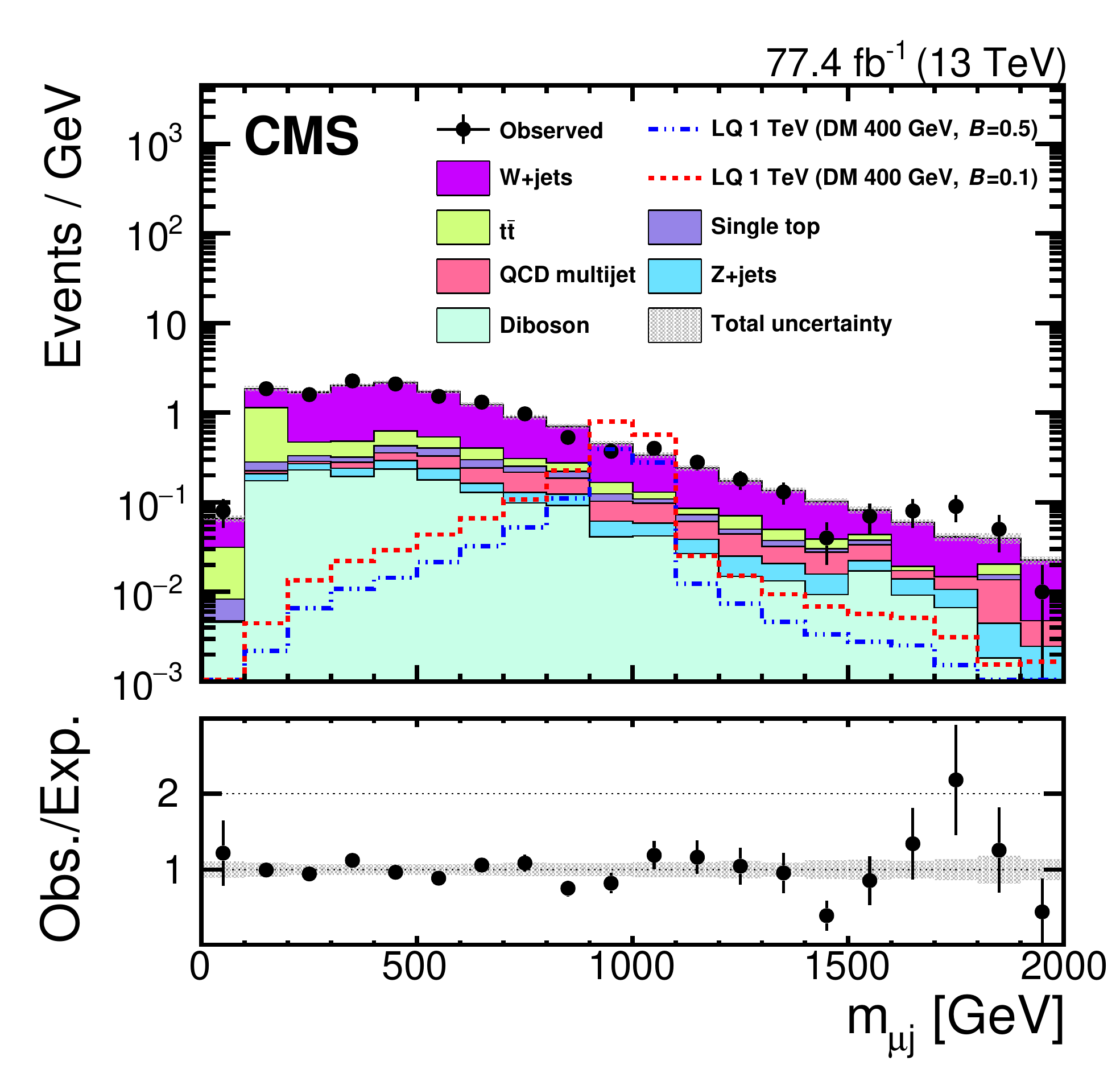}
\caption{
The observed distribution of \Mmuj in comparison to the
post-fit SM background predictions for the combined 2016
and 2017 data sets.
``Post-fit'' means that the constraints from
the maximum likelihood fit are incorporated.
The unstacked predictions for two signal models
with $\MLQ=1000\GeV$ and $\mdm=400\GeV$ are also shown:
one with $\bzero=0.5$ and the other with $\bzero=0.1$.
The difference is just an overall relative normalization of about 2 for the latter compared to the former.
The ratio of the observed results to the total SM prediction
is shown in the lower panel.
The vertical error bars on the data points are statistical.
The gray band shows the total uncertainty in the background prediction,
including both statistical and systematic terms.
}
\label{fig:STplotsLQLog}
\end{figure}

The observed distribution of \Mmuj is presented
in Fig.~\ref{fig:STplotsLQLog}.
The results are shown in comparison to the
post-fit predictions for the SM background, where
``post-fit'' means that the constraints from
the maximum likelihood fit are incorporated.
For purposes of illustration,
the predictions of two signal models
with $\MLQ=1000\GeV$ and $\mdm=400\GeV$ are also shown:
one with $\bzero=0.5$ and the other with $\bzero=0.1$.
The difference is just an overall relative normalization of about 2 for the latter compared to the former.
Numerical values are given in Table~\ref{tab:results}.

The data are found to be consistent with the SM predictions
within the uncertainties.
There is a small excess of events above the SM prediction
in the \Mmuj region between 1600 and 1900\GeV,
consistent with a statistical fluctuation.
This excess corresponds to a statistical significance of around 1.5~standard deviations.
Thus, we do not obtain evidence for DM or LQ production.

\begin{table*}[tbh]
\topcaption{
Observed number of events, post-fit SM background predictions and post-fit uncertainties
for the combined 2016 and 2017 data sets.
``Electroweak" refers to the sum of expected events
from the single top quark, \cPZ\ boson, and diboson background processes.
The predictions for two signal models with
$\MLQ=1000\GeV$ and $\mdm=400\GeV$ are also shown:
one with $\bzero=0.5$ and the other with $\bzero=0.1$.
The uncertainties represent the statistical
and systematic terms added in quadrature.
}
\centering
\begin{tabular}{lc}
\hline
Process &  Events \\
\hline
\wjets                       & $911 \pm 55$ \\
\ttbar                       & $185\pm25$   \\
Electroweak                  & $241\pm26$   \\
QCD                          & $63\pm26$    \\[\cmsTabSkip]
Total SM background          & $1401\pm 40$     \\[\cmsTabSkip]
Signal; $\bzero=0.5$, $\MLQ=1000\GeV$, $\mdm=400\GeV$  & $96\pm 8$ \\
Signal; $\bzero=0.1$, $\MLQ=1000\GeV$, $\mdm=400\GeV$  & $195\pm 16$ \\[\cmsTabSkip]
Observed                     & 1390 \\
\hline
\end{tabular}
\label{tab:results}
\end{table*}

A binned maximum likelihood fit is applied to the
\Mmuj distribution in the signal region.
The fitted parameters are the yields of the individual
background components listed in Table~\ref{tab:results},
the signal yield,
and various nuisance parameters.
The nuisance parameters are introduced to treat systematic uncertainties.
Log-normal probability distributions are used for nuisance parameters
that affect the normalizations of the signal and background yields.
Gaussian probability distributions are used for nuisance parameters
that affect the shape of the \Mmuj distribution.
All normalization uncertainties are taken to be fully correlated across bins
except those that are statistical in origin,
which are assumed to be uncorrelated.
The overall postfit uncertainty in the dominant \wjets background is substantially smaller than the prefit value shown in Table~\ref{tab:systematics} because the normalization and shape of this background is highly constrained by the lower side of the mass distribution.

Upper limits at 95\% confidence level (\CL) are determined
on the product
of the signal production cross section and branching fraction.
These limits are calculated using
a modified frequentist approach with the \CLs criterion~\cite{CLS2,Read:2002hq}
and an asymptotic approximation for the test
statistic~\cite{Cowan:2010js,CMS-NOTE-2011-005}.
The limits are determined as a function of \MLQ and~\mdm.

\begin{figure}[tbh!]
\centering
\includegraphics[width=0.49\textwidth]{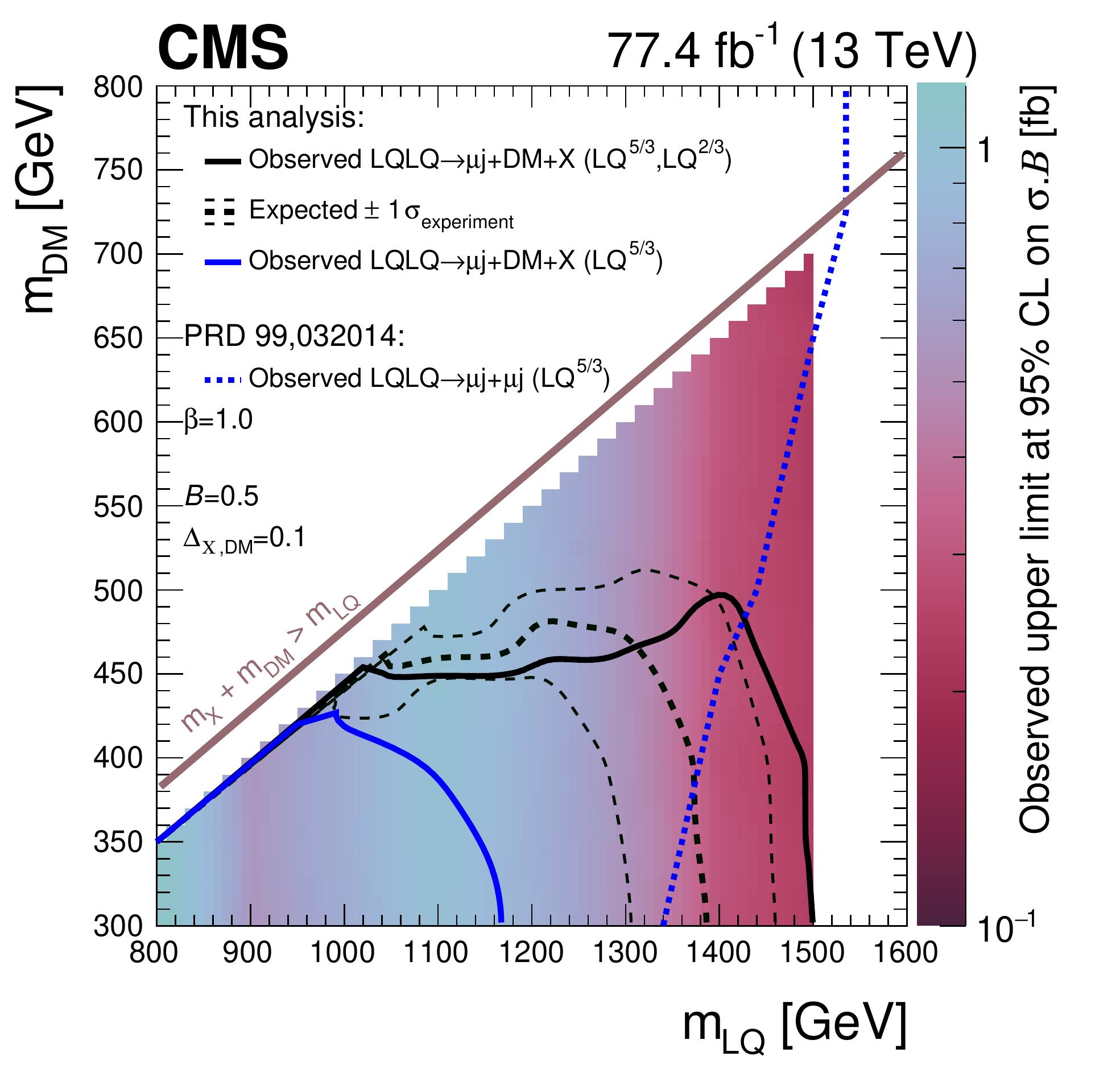}
\includegraphics[width=0.49\textwidth]{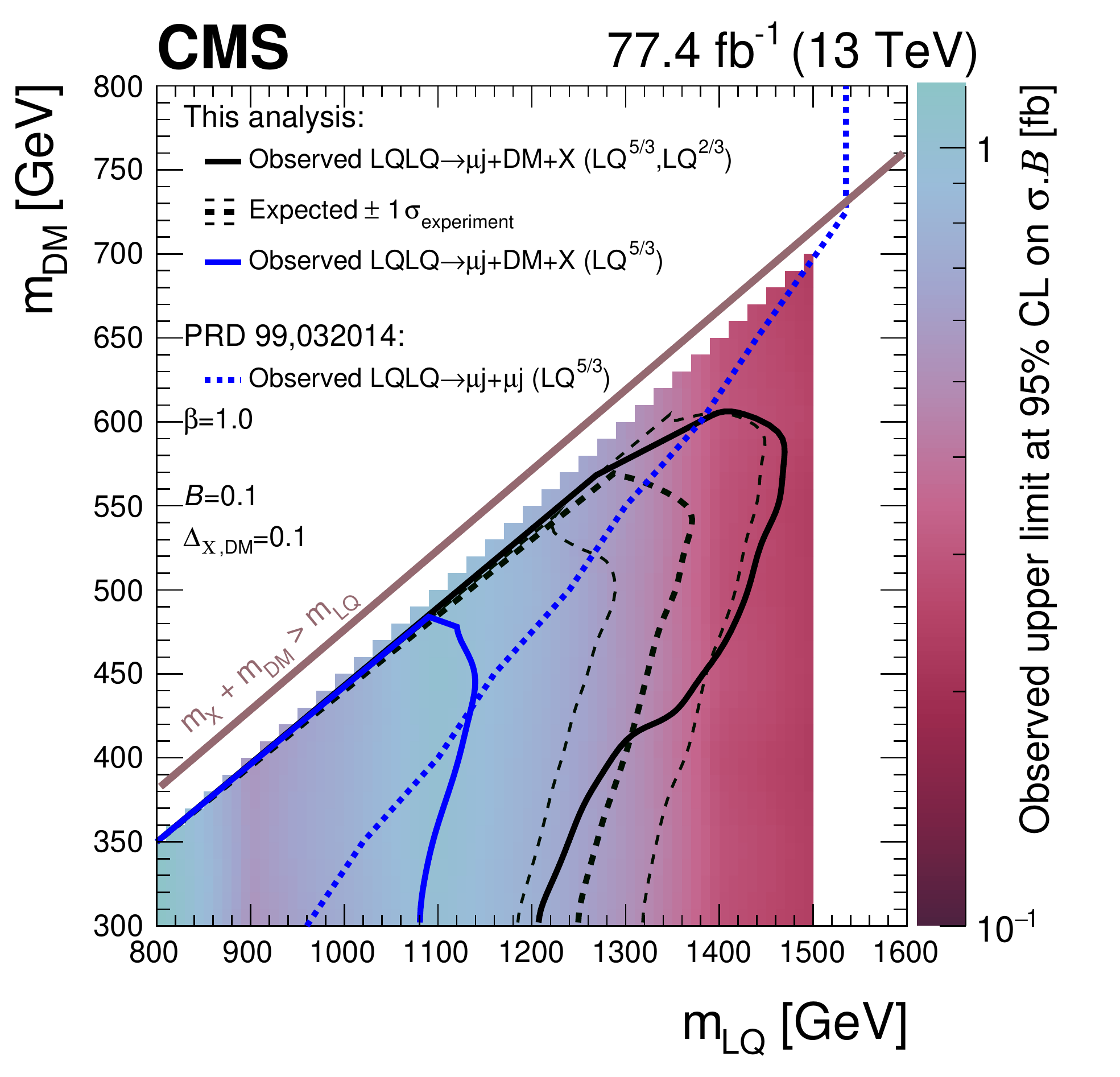}
\caption{
Observed 95\% \CL upper limits on the product of cross section and
branching fraction
for the signal model of Fig.~\ref{fig:feynman}
assuming  $\bzero = \left.\lqbf\right|_{\mdm=\mx=0}$  to be
(\cmsLeft) 0.5
or
(\cmsRight)~0.1.
The solid and dashed black curves show the observed and expected
95\% \CL exclusion curves, taking into account both upper and lower components of the LQ doublet.
The solid blue curve shows the observed exclusion limit for the upper component of the LQ doublet, i.e. to a muon and a \cPqc\ quark.
The dotted blue curve
shows the corresponding observed limits from the recast
of the results from a search for
pair produced second-generation LQs~\cite{Sirunyan:2018ryt}.
}
\label{fig:LimitplotsLQ}
\end{figure}

Figure~\ref{fig:LimitplotsLQ} (\cmsLeft) presents the results for $\bzero=0.5$. The region to the right and below the diagonal line is where LQ decay to DM and \X is allowed. The solid and dashed black curves show the observed and expected exclusion limits, respectively, taking into account contributions from both upper and lower components of the LQ doublet. Combinations of \MLQ and \mdm to the left and below the solid curve are excluded. It can be seen that the signal scenario of Fig.~\ref{fig:feynman} is excluded for values of  \mdm up to 500\GeV for $\MLQ\approx1400\GeV$ and up to 300\GeV for $\MLQ\approx1500\GeV$.

 This is the first test of the co-annihilation process proposed in~\cite{Baker:2015qna}, thus it is not possible to make a direct comparison with other results. To give an indication of the extent to which our results explore untested regions of the \MLQ-\mdm parameter space we have therefore performed a recast of the results from a search for pair produced second-generation LQs, each decaying to a muon and a \cPqc\ quark~\cite{Sirunyan:2018ryt}.
 The recast is performed in two steps. In the first step, the change in the branching fraction of the LQ to a lepton and a quark is calculated once the decay of the LQ to DM and \X is allowed. The change depends on the parameters of the model including \dmx, \bzero, LQ and DM mass~\cite{Baker:2015qna}. The altered branching fraction is then used to find the exclusion contour in the plane of \MLQ and \mdm.
The resultant limit contour is shown as the dotted blue curve. This limit contour may be compared with the solid blue curve, which shows the observed exclusion limit that would be obtained from the present analysis if only the upper component of the LQ doublet, decaying to a muon and a \cPqc\ quark, contributed to the potential signal.

 Figure~\ref{fig:LimitplotsLQ} (\cmsRight) shows results obtained assuming the somewhat smaller value of 0.1 for \bzero. In this case, the upper limit of excluded DM mass is extended significantly, reaching a maximum of $\approx$600\GeV for $\MLQ \approx$1400\GeV.

\section{Summary}
\label{sec:summary}

A search has been performed for dark matter in events containing a muon,
a jet, and significant missing transverse momentum.
The study is conducted using proton-proton collision data at $\sqrt{s} = 13\TeV$
recorded with the CMS detector,
corresponding to an integrated luminosity of 77.4\fbinv.
It is assumed that dark matter is produced through the production
of a leptoquark pair,
with one leptoquark decaying to a muon and a jet,
and the other to dark matter and low-\pt standard model particles.
The analysis is performed by searching for a peak
in the leptoquark candidate invariant mass \Mmuj distribution formed
from the highest \pt muon and jet in an event,
with the requirement of significant missing transverse momentum,
as is expected from the presence of dark matter.
The observation of such a peak in this novel search would provide
strong evidence for the existence of both dark matter particles and leptoquarks.
The data are observed to agree with the standard model background
predictions within the uncertainties.
Upper limits on the product of the cross section and branching fraction
are obtained at 95\% confidence level as a function of
the leptoquark and dark matter particle masses.
For the first benchmark scenario considered, dark matter masses up to 500\GeV
are excluded for leptoquark masses $\MLQ\approx1400\GeV$,
and up to 300\GeV for $\MLQ\approx1500\GeV$.	
For the second benchmark scenario, dark matter masses up to 600\GeV are excluded for $\MLQ\approx 1400\GeV$.

\begin{acknowledgments}

We congratulate our colleagues in the CERN accelerator departments for the excellent performance of the LHC and thank the technical and administrative staffs at CERN and at other CMS institutes for their contributions to the success of the CMS effort. In addition, we gratefully acknowledge the computing centers and personnel of the Worldwide LHC Computing Grid for delivering so effectively the computing infrastructure essential to our analyses. Finally, we acknowledge the enduring support for the construction and operation of the LHC and the CMS detector provided by the following funding agencies: BMWFW and FWF (Austria); FNRS and FWO (Belgium); CNPq, CAPES, FAPERJ, FAPERGS, and FAPESP (Brazil); MES (Bulgaria); CERN; CAS, MoST, and NSFC (China); COLCIENCIAS (Colombia); MSES and CSF (Croatia); RPF (Cyprus); SENESCYT (Ecuador); MoER, ERC IUT, and ERDF (Estonia); Academy of Finland, MEC, and HIP (Finland); CEA and CNRS/IN2P3 (France); BMBF, DFG, and HGF (Germany); GSRT (Greece); NKFIA (Hungary); DAE and DST (India); IPM (Iran); SFI (Ireland); INFN (Italy); MSIP and NRF (Republic of Korea); LAS (Lithuania); MOE and UM (Malaysia); BUAP, CINVESTAV, CONACYT, LNS, SEP, and UASLP-FAI (Mexico); MBIE (New Zealand); PAEC (Pakistan); MSHE and NSC (Poland); FCT (Portugal); JINR (Dubna); MON, RosAtom, RAS, RFBR, and NRC KI (Russia); MESTD (Serbia); SEIDI, CPAN, PCTI, and FEDER (Spain); Swiss Funding Agencies (Switzerland); MST (Taipei); ThEPCenter, IPST, STAR, and NSTDA (Thailand); TUBITAK and TAEK (Turkey); NASU and SFFR (Ukraine); STFC (United Kingdom); DOE and NSF (USA).

 \hyphenation{Rachada-pisek} Individuals have received support from the Marie-Curie programme and the European Research Council and Horizon 2020 Grant, contract No. 675440 (European Union); the Leventis Foundation; the A. P. Sloan Foundation; the Alexander von Humboldt Foundation; the Belgian Federal Science Policy Office; the Fonds pour la Formation \`a la Recherche dans l'Industrie et dans l'Agriculture (FRIA-Belgium); the Agentschap voor Innovatie door Wetenschap en Technologie (IWT-Belgium); the F.R.S.-FNRS and FWO (Belgium) under the ``Excellence of Science - EOS" - be.h project n. 30820817; the Ministry of Education, Youth and Sports (MEYS) of the Czech Republic; the Lend\"ulet (``Momentum") Programme and the J\'anos Bolyai Research Scholarship of the Hungarian Academy of Sciences, the New National Excellence Program \'UNKP, the NKFIA research grants 123842, 123959, 124845, 124850 and 125105 (Hungary); the Council of Science and Industrial Research, India; the HOMING PLUS programme of the Foundation for Polish Science, cofinanced from European Union, Regional Development Fund, the Mobility Plus programme of the Ministry of Science and Higher Education, the National Science Center (Poland), contracts Harmonia 2014/14/M/ST2/00428, Opus 2014/13/B/ST2/02543, 2014/15/B/ST2/03998, and 2015/19/B/ST2/02861, Sonata-bis 2012/07/E/ST2/01406; the National Priorities Research Program by Qatar National Research Fund; the Programa Estatal de Fomento de la Investigaci{\'o}n Cient{\'i}fica y T{\'e}cnica de Excelencia Mar\'{\i}a de Maeztu, grant MDM-2015-0509 and the Programa Severo Ochoa del Principado de Asturias; the Thalis and Aristeia programmes cofinanced by EU-ESF and the Greek NSRF; the Rachadapisek Sompot Fund for Postdoctoral Fellowship, Chulalongkorn University and the Chulalongkorn Academic into Its 2nd Century Project Advancement Project (Thailand); the Welch Foundation, contract C-1845; and the Weston Havens Foundation (USA).

\end{acknowledgments}

\bibliography{auto_generated}
\cleardoublepage \appendix\section{The CMS Collaboration \label{app:collab}}\begin{sloppypar}\hyphenpenalty=5000\widowpenalty=500\clubpenalty=5000\vskip\cmsinstskip
\textbf{Yerevan Physics Institute, Yerevan, Armenia}\\*[0pt]
A.M.~Sirunyan, A.~Tumasyan
\vskip\cmsinstskip
\textbf{Institut f\"{u}r Hochenergiephysik, Wien, Austria}\\*[0pt]
W.~Adam, F.~Ambrogi, E.~Asilar, T.~Bergauer, J.~Brandstetter, M.~Dragicevic, J.~Er\"{o}, A.~Escalante~Del~Valle, M.~Flechl, R.~Fr\"{u}hwirth\cmsAuthorMark{1}, V.M.~Ghete, J.~Hrubec, M.~Jeitler\cmsAuthorMark{1}, N.~Krammer, I.~Kr\"{a}tschmer, D.~Liko, T.~Madlener, I.~Mikulec, N.~Rad, H.~Rohringer, J.~Schieck\cmsAuthorMark{1}, R.~Sch\"{o}fbeck, M.~Spanring, D.~Spitzbart, A.~Taurok, W.~Waltenberger, J.~Wittmann, C.-E.~Wulz\cmsAuthorMark{1}, M.~Zarucki
% ...
\vskip\cmsinstskip
% ...
1:  Also at Vienna University of Technology, Vienna, Austria\\
% ...
\end{sloppypar}
\end{document}